\documentclass{sig-alt-hotnets}

\usepackage[font=bf]{caption}
\newcommand{\subparagraph}{}
\usepackage[compact]{titlesec}
\usepackage{times}
\usepackage{subfig}
\usepackage{amsmath}
\usepackage{subfig}
\usepackage{epsfig,endnotes,color,paralist, multirow,appendix}
\usepackage{times}
\usepackage{xspace}
\usepackage{url}

\newcounter{packednmbr}

\newenvironment{packeditemize}{
\begin{list}{$\bullet$}{
\setlength{\itemsep}{1.5pt}
\setlength{\labelwidth}{8pt}
\setlength{\leftmargin}{10pt}
\setlength{\labelsep}{3pt}
\setlength{\listparindent}{\parindent}
\setlength{\parsep}{1.5pt}
\setlength{\parskip}{1.5pt}
\setlength{\topsep}{1.5pt}}}{\end{list}}

\newcommand{\tightcaption}[1]{\vspace{-8pt}\caption{#1}\vspace{-8pt}}

\newcommand{\comment}[1]{}

\usepackage{footnote}
\usepackage{pifont}

\newcommand{\mypara}[1]{\medskip\noindent{\bf {#1}:}~}
\newcommand{\myparatight}[1]{\smallskip\noindent{\bf {#1}:}~}

\title{A Framework to Quantify the Benefits of Network Functions
Virtualization in Cellular Networks}

\author{Zafar A. Qazi$^\dag$, Vyas Sekar$^\star$, Samir R. Das$^\dag$ \\ \normalsize{$^\dag$ Stony Brook University $^\star$ Carnegie Mellon}}

\begin{document}
\maketitle

 \newcommand{\itembold} [1] {\item{\bf#1}}
 \newcommand{\Vector}[1]{<#1>}
 \newcommand{\TrafficClass}{\ensuremath{c}} 
 \newcommand{\Epoch}{\ensuremath{\mathit{e}}}
 \newcommand{\TrafficVolume}{\ensuremath{|T_{\TrafficClass,\Epoch}|}\xspace} 
 \newcommand{\Set}[2]{\ensuremath{#1=(#2)}\xspace}
 \newcommand{\Class}{\ensuremath{\mathit{c}}}

 \newcommand{\MiddleboxFunction}[1]{\ensuremath{\NF_{#1}}}
 \newcommand{\MiddleboxFunctionIndex}{\ensuremath{\mathit{m}}}
 \newcommand{\MiddleboxFunctionIndexGrave}{\ensuremath{\MiddleboxFunctionIndex'}}
 \newcommand{\MiddleboxFunctionIndexGraveGrave}{\ensuremath{\MiddleboxFunctionIndex^{''}}}

 \newcommand{\ResourceIndex}{\ensuremath{\mathit{r}}}
 \newcommand{\Equation}[1]{\begin{equation} {#1} \end{equation}}
 \newcommand{\Type}{\ensuremath{\mathit{t}}}
 
\newcommand{\PlatformType}{\ensuremath{P_{\Location,\Type}}}
 \newcommand{\Location}{\ensuremath{\mathit{l}}}
 \newcommand{\LocationGrave}{\ensuremath{\mathit{\grave{l}}}}
 \newcommand{\Platform}{\ensuremath{p}}
 \newcommand{\PlatformGrave}{\ensuremath{\Platform'}}
 
\newcommand{\Capacity}{\ensuremath{\mathit{Cap}_{\Location,\Type,\MiddleboxIndex}}}
 
\newcommand{\FixedMBCost}{\ensuremath{{\mathit{Fixed_{\Location,\Type}}}}}
 \newcommand{\FixedMBCostPlatform}{\ensuremath{{\mathit{Fixed_{\Location(\Platform),\Type(\Platform)}}}\xspace}}
 \newcommand{\VariableMBCost}{\ensuremath{{\mathit{Var_{\Location,\Type}}}\xspace}}
 \newcommand{\VariableMBCostPlatform}{\ensuremath{{\mathit{Var_{\Location(\Platform),\Type(\Platform)}}}\xspace}}
 \newcommand{\ElasticMBCost}{\ensuremath{{\mathit{Elas_{\Location,\Type}}}\xspace}}
 \newcommand{\ElasticMBCostPlatform}{\ensuremath{{\mathit{Elas_{\Location(\Platform),\Type(\Platform)}}}\xspace}}
 \newcommand{\ActiveMB}{\ensuremath{Active_{\Location,\Type,\MiddleboxIndex}}}

 \newcommand{\FootPrint}{\ensuremath{{\mathit{FP}_{\TrafficClass,\MiddleboxFunctionIndex,\Type}}\xspace}}  
 \newcommand{\FootPrintPlatform}{\ensuremath{{\mathit{FP}_{\TrafficClass,\MiddleboxFunctionIndex,\Type(\Platform)}}\xspace}}  
 \newcommand{\Resource}{\ensuremath{\mathit{Res}_{\Location,\Type,\MiddleboxIndex}}}
 \newcommand{\ResourceEpoch}{\ensuremath{{\mathit{Res_{\Location,\Type,\MiddleboxIndex, \Epoch}}}\xspace}} 
 \newcommand{\Load}{\ensuremath{{\mathit{Load_{\Location,\Type,\MiddleboxIndex,\Epoch}}}\xspace}} 

  \newcommand{\MiddleboxInstance}[1]{\ensuremath{\Platform_{#1}}}
 \newcommand{\MiddleboxInstanceB}[1]{\ensuremath{\PlatformGrave_{#1}}}
   \newcommand{\NetworkCostInstance}{\ensuremath{\mathit{Lat}_{ \Platform_{\MiddleboxFunctionIndex}, \PlatformGrave_{\MiddleboxFunctionIndexGrave}} }}
  \newcommand{\CapPerPlatform}{\ensuremath{Cap_{\Platform}}} 
   \newcommand{\CapPerFunction}{\ensuremath{Cap_{\Platform_{\MiddleboxFunctionIndex}}}}

      \newcommand{\Fraction}{\ensuremath{f}}
      \newcommand{\FractionOfTraffic}{\ensuremath{\Fraction_{\TrafficClass,\Epoch, \Platform_{\MiddleboxFunctionIndex}, \PlatformGrave_{\MiddleboxFunctionIndexGrave}}}}
      \newcommand{\FractionOfTrafficSingle}{\ensuremath{\Fraction_{\TrafficClass,\Epoch, \Platform_{\MiddleboxFunctionIndex}}}}

     \newcommand{\ResourcePerPlatform}{\ensuremath{Res_{\Platform}}}
     \newcommand{\ResourcePerPlatformPerEpoch}{\ensuremath{Res_{\Platform,\Epoch}}}
        \newcommand{\LoadPerFunction}{\ensuremath{\mathit{LoadPerNF}_{\Platform_{\MiddleboxFunctionIndex},\Epoch}}}
       \newcommand{\PolicyChain}{\ensuremath{\mathit{SC}_{\TrafficClass}}}
        \newcommand{\LoadPerPlatform}{\ensuremath{Load_{\Platform,\Epoch}}}
           \newcommand{\ActivePlatform}{\ensuremath{Active_{\Platform}}}
          \newcommand{\PolicyChainSize}{\ensuremath{|\PolicyChain|}} 
	 \newcommand{\ThresholdPerChain}{\ensuremath{\mathit{PC}_{\Class}}} 
     	 
\newcommand{\Flexible}{\emph{FlexHW}\xspace} 
\newcommand{\Dedicated}{\emph{Single}\xspace} 
\newcommand{\Cloud}{\emph{Cloud}\xspace} 
\newcommand{\Eqref}[1]{Eq~\eqref{#1}}

\newcommand{\NF}{NF\xspace} 
\newcommand{\NFs}{NFs\xspace}

\section*{Abstract}
Network functions virtualization (NFV) is an appealing  vision that  promises
to dramatically reduce capital and operating expenses for cellular providers.
 However, existing efforts in this space leave open broad
issues about how NFV deployments should be instantiated or how they should be
provisioned. In this paper, we present an initial attempt at a  framework that
will help network operators systematically evaluate the potential benefits that
different points in the NFV design space can offer.

\section{Introduction}

  Cellular networks today incur high capital costs in deploying a broad range of
expensive and inflexible hardware appliances which include both
cellular-specific functions such as Serving and Packet Data Networks Gateway (S/P-GW), IP Multimedia System (IMS) elements like 
Call Session Control Functions (CSCFs)~\cite{lte}, as well as more traditional
network appliances (e.g., NATs, firewalls, proxies)~\cite{morleysigcomm}.
Furthemore, they incur high management complexity on many fronts: diverse
protocols and standards (e.g., 3G, 4G), multiple types of services (e.g.,
video, voice, messaging),  and fine-grained policy requirements (e.g.,
per-user accounting of data and voice calls and specialized video services).

In conjunction, these effects have led to a state where cellular providers face
an uphill battle with the trend in the gap  between  revenues and their capital
and operating expenses  being quite unfavorable~\cite{verizononstalk}.
Furthermore, current deployments are inflexible on several accounts: they
cannot react to changing demands and policies and  the timescales of innovation
are hindered by vendor support.

\begin{figure}[t]
 \centering
\includegraphics[width=220pt]{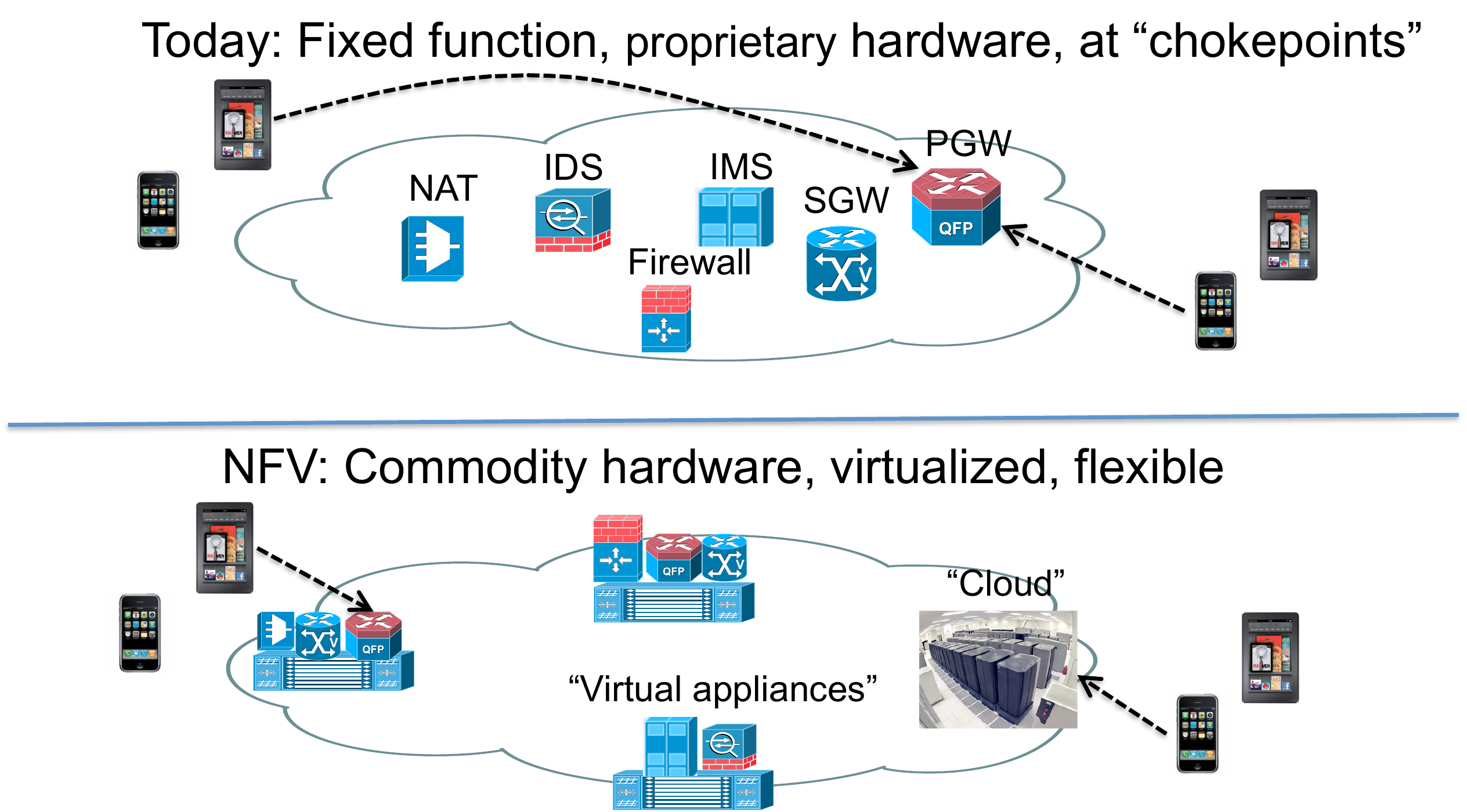}
\tightcaption{The current fixed and proprietary implementation of network functions \textit{vs} the NFV vision of elastic, cost effective, mix-match and potentially hybrid deployment of network functions. }
\label{fig:nfv_vision}
\end{figure}

This trend is likely unsustainable in the longer term and has motivated the
case for  \emph{network functions virtualization} (NFV)~\cite{nfv_whitepaper}. The  motivation in NFV is to bring the benefits that cloud computing has
provided for the IT industry to network operators: accelerating the pace of
innovation by reducing the cycles to deploy new equipment, economies of scale
provided by commodity hardware, resource multiplexing via virtualization,
dynamic provisioning and elastic scaling, and the ability to experiment with
new services without significant upfront costs~\cite{nfv_whitepaper}(see also Figure~\ref{fig:nfv_vision}).  The
high-level idea is that the various {\em network functions} (\NFs) for
cellular-, IP-, and application-specific services  can be replaced by {\em
virtualized}  applications on commodity hardware platforms. 


As highlighted in an early whitepaper~\cite{nfv_whitepaper}, this vision builds
upon, and is complementary, to existing work in the  software-defined
networking (SDN). The main differences are that: (a)  it broadens the scope of
the data plane functions beyond OpenFlow-enabled switches; and (b) it focuses
more on the carriers and the services they would like to offer to their
customers.


While the above long-term vision is appealing, it also leaves 
 open several aspects of the design space that operators would need
 to address in practice w.r.t.: 
\begin{packeditemize} 
\item the
type of {\em platforms} (e.g., pure cloud, or fixed-but-flexible hardware
infrastructure);
\item  the type of {\em provisioning} (i.e., where to place new flexible 
 hardware or datacenters); and 
\item {\em demand distribution} (i.e., how to route user demands to different function instances to meet logical policy requirements).
\end{packeditemize}

 Given this broad design space, there are  several possible NFV instantiations
by combining choices across the individual platform, provisioning, and
distribution dimensions.   For instance, we can imagine a consolidated  
 deployment where the provider has a small number of datacenters at which the
network functions (\NFs) can be run in a dynamic, elastic manner. At the other
extreme, we can imagine current deployments with specialized NF hardware. 
 We can also envision ``nano-datacenter'' like models with flexible \NF 
 hardware distributed throughout the network~\cite{nanodc}. 
 These deployments will naturally have different provisioning, operational, and
performance characteristics.

Given this diverse and broad design space, network operators will need
systematic decision systems to help them evaluate the cost-benefit
tradeoffs of different  points in the design space. We highlight some
motivating scenarios in \S\ref{sec:motivation}.  Furthermore, even before
embarking on rearchitecting their network infrastructure~\cite{attdomain2.0},
operators need to first  {\em quantify} the potential CAPEX and OPEX benefits
that  specific NFV strateges might offer.

This paper is a first attempt to shed light on these issues. To this end, we
cast  the NFV deployment  problem as a systematic optimization framework (\S\ref{sec:model}). Our
framework is general enough to capture different points in the design space and
also consider hybrid NFV different deployments (e.g., some combination of pure
cloud and fixed hardware).  In order to estimate the potential benefits of
different NFV strategies, the operators provide  as input historical traffic
demands and policy based service chaining requirements for different traffic
patterns (\S\ref{sec:inputs}). Our  framework will then output guidelines on the optimal
provisioning  strategy and the cost benefits it offers.  Operators can use such
a framework for ``what-if'' analysis to  evaluate the cost-benefit tradeoffs of
different deployment  strategies.

As illustrative examples we show how
operators can use our framework to  evaluate the benefits of different NFV
designs (\S\ref{sec:usecase}). For instance, we observe that using flexible hardware minimizes the deployment cost in many scenarios. We also conduct a sensitivity analysis to evaluate the effects of changing different input parameters on the optimal deployment strategy.



\section{Motivation}
\label{sec:motivation}
 We begin by outlining  the  
 NFV design space and then use  motivating scenarios to 
 highlight how the optimal NFV strategy  depends 
 on the workload and cost factors.



\subsection{Design Space of Cellular NFV}
\label{sec:motivation:designspace}
We identify three key dimensions for the design space:
\begin{packeditemize}
\item {\em Platform type:} Network functions (\NFs) can  be realized in
 many ways. Today, each NF is  a dedicated appliance
(\Dedicated) providing a specialized capability. Going forward, one can imagine
a flexible commodity hardware (\Flexible) that can be repurposed to run
different types of NFs on demand~\cite{clickos,comb,xomb}.  Going one step
further, we can imagine that the functions are themselves outsourced to a
\Cloud service that can  elastically scale resources for different
NFs.\footnote{Industry reports
use the terms \Flexible and \Cloud rather loosely. They are however quite distinct in their cost-performance
 tradeoffs and thus one of our goals in formalizing the NFV problem space is to
crystallize these loose characterizations of NFV instantiations.}  
\item {\em Provisioning and placement:} A key operational decision is deciding
how and where to provision  \NF platforms. At one extreme, the
provider can choose a single \Cloud location. At  the other extreme we can
envision a nano-datacenter model where every cell base station has an
associated mini-\Cloud (e.g.,~\cite{nanodc}). We can also consider simple hybrids where each location
has  pre-provisioned \Dedicated and \Flexible boxes and we have a few
 \Cloud locations.  

\item {\em Scaling and distribution strategies:}  Given 
 a specific provisioning/placement strategy, another aspect of the 
design space is how the available hardware resources (possibly elastic)
 are used to serve the (varying) offered load. 
 Again, we can envision 
 several possible strategies here: optimal load balancing, 
 or  routing to nearest available instance of a specific 
  \NF, or offloading to the cloud beyond 
a threshold value of load.

%


\end{packeditemize}

\subsection{Motivating Scenarios}


\begin{figure}[t]
 \centering
\includegraphics[width=220pt]{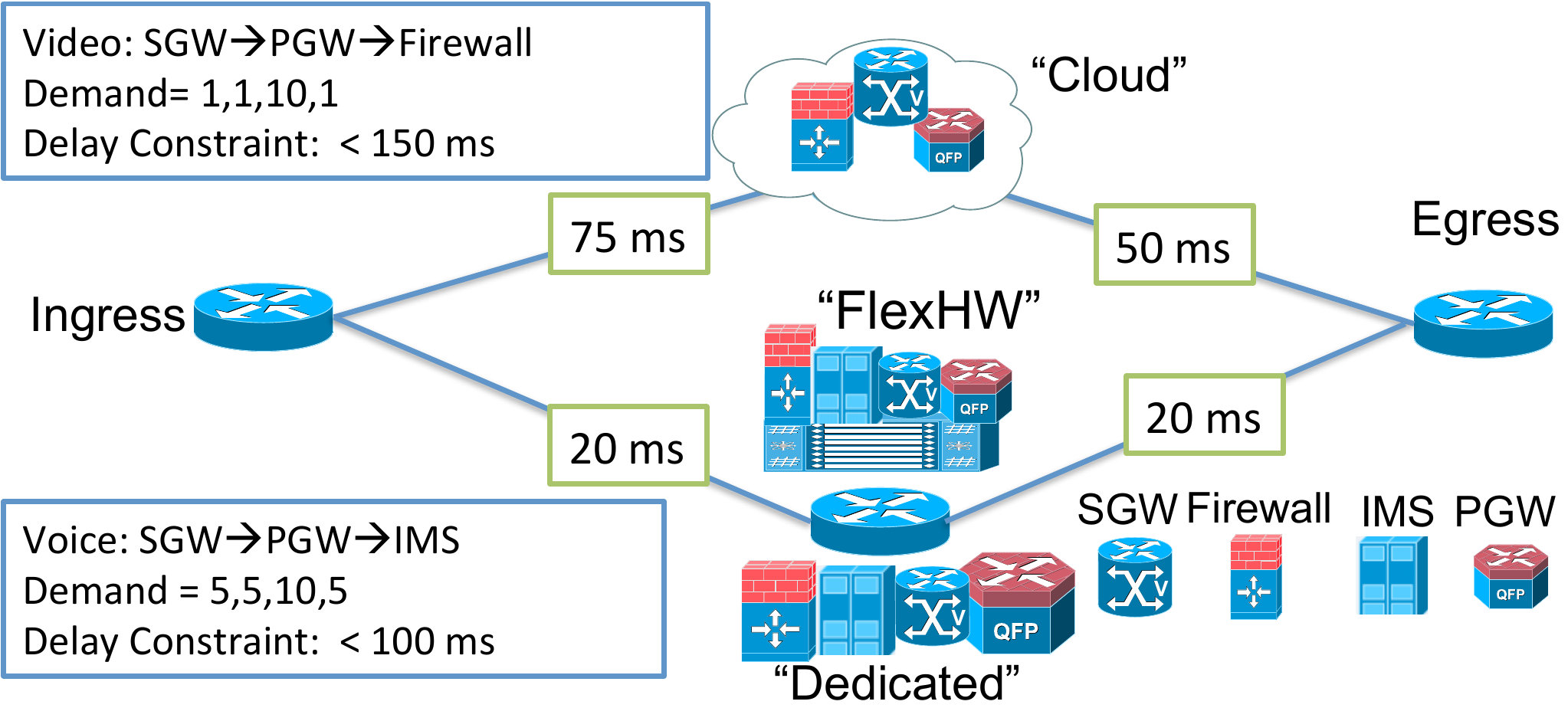}
\tightcaption{Example to motivate the different design tradeoffs in provisioning.}
\label{fig:motivation}
\end{figure}

\begin{figure}[t]
 \centering
\includegraphics[width=220pt]{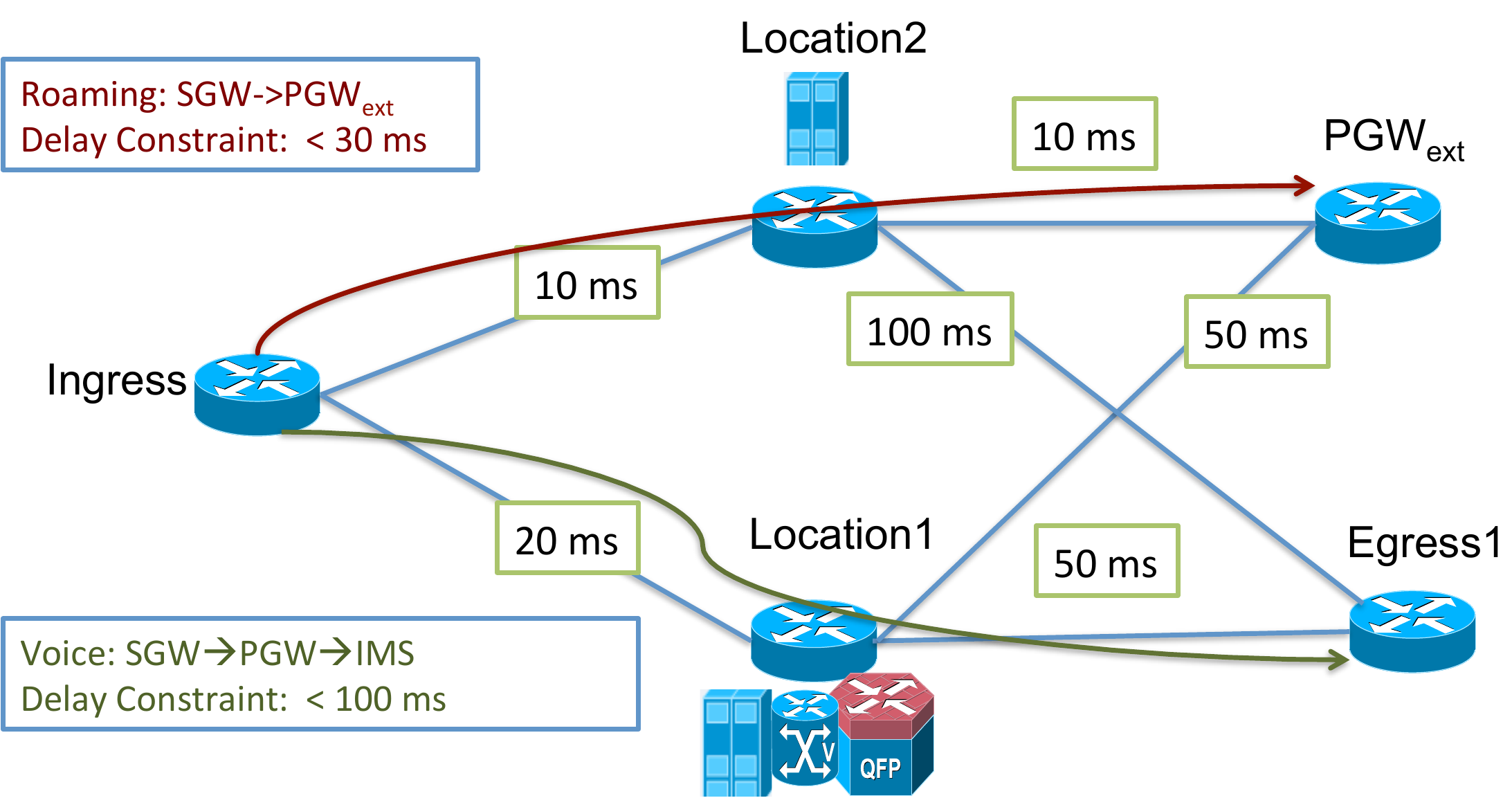}
\tightcaption{Example to illustrate the different design tradeoffs in functional placement and routing.}
\label{fig:motivation2}
\end{figure}

We use the simple scenarios in Figure~\ref{fig:motivation} and Figure~\ref{fig:motivation2}
 to illustrate how different points in the NFV design space may be optimal
depending on the performance constraints, traffic patterns, policy chains and CAPEX and OPEX.

\mypara{Optimizing provisioning cost} Consider the single traffic class, {\em Video}, with 
 traffic entering at ``Ingress'' and exiting at ``Egress'' with the {\em service chain} SGW-PGW-FIREWALL. Suppose the traffic volume for this traffic class across four time epochs is \{1,1,10,1\}. 
Assume that the (fixed) cost per unit capacity for fixed in-network provisioning is 20. 
On the other hand, if we use \Cloud to dynamically provision the needed resource,
assume the amortized cost per unit capacity provisioned is 10. 
With fixed pre-provisioning we need to allocate for peak traffic and thus
 a capacity of 10 units costing 200. But when outsourcing the processing
to the cloud, the cost is only 130.  Thus, we decide to provision the entire
service chain for {\em Video} in the \Cloud. 

%

\mypara{Performance constraint}
Now, assume that we add another traffic class, {\em Voice}, with the {\em service chain} SGW-PGW-IMS. 
Assume that  {\em Voice} has a performance constraint that the average 
latency should be less than 100~ms. Then,  we cannot use the \Cloud for processing traffic belonging to this class. 
Assume also that the traffic volume for {\em Voice} across the
 four time epochs is \{5,5,10,5\}. 
Now, a hybrid solution where we provision a capacity of 10 units on fixed hardware and 
10 units on the \Cloud dynamically only in epoch 3 will be the most cost effective. This solution has 
cost 300, while provisioning entirely on the \Cloud costs 380 (if there are no performance constraints)  and provisioning
entirely on a fixed hardware costs 400. 




\mypara{Functional placement and routing}
To illustrate placement/routing issues consider the network in Figure~\ref{fig:motivation2}.
We have two traffic classes {\em Roaming} and {\em Voice}. The traffic belonging to  {\em Roaming}  
requires processing SGW-$\mathrm{PGW}_{ext}$, where $\mathrm{PGW}_{ext}$ is the PGW of the external network to which the traffic belongs. On the other hand, {\em Voice} needs processing using
the service chain SGW-PGW-IMS. Assume that the carrier normally prefers to provision
all resources in ``Location 1''  due to cost or management issues.  However,
when the {\em Roaming} traffic enters the network, it must move the related SGW
processing to ``Location 2'' as going via ``Location 1'' violates the delay constraint for 
{\em Roaming}. Assume also after a network upgrade the
delay on the  Location~1-$\mathrm{PGW}_{ext}$ link falls to 5\,ms. Then, the SGW processing
can move back to Location 1.  Moving the SGW processing
back and forth depending on traffic
and network changes could be facilitated via use of \Flexible.

%


The above  scenarios highlights the tradeoffs and considerations
operators need to make.  It is important for operators to be able to
analyze these tradeoffs before rolling out new  NFV deployments.  However, given the size and complexity of modern cellular deployments, 
it may be  impractical 
for operators to manually evaluate the entire space of design
options. Thus, our  goal is to develop a decision support framework that 
allows cellular operators to systematically explore different design options
before deploying new hardware.  For instance,  they would
like to specify policy requirements (e.g., service chains for different user
classes) and performance constraints (e.g., load, congestion, or latency) and use such
a decision support system to choose the right mix of platform, provisioning,
and distribution strategies from the broader NFV design space highlighted
above.


\section{Inputs and Requirements}
\label{sec:inputs}

We begin by describing the requirements and inputs that 
 operators need to provide to our framework. This data can be obtained 
 from their network logs, policy configurations, and  vendor-specific 
 benchmarks.

\myparatight{Traffic patterns} Cellular traffic is divided into different
logical {\em classes}  based on  different user/customer demands.  For
 example, the classes may capture  {\em regular users} vs.\ {\em roaming
customers} vs.\ {\em machine-to-machine} (M2M) traffic, with different
requirements.  We assume that the operator has historical demand patterns, with
$\TrafficVolume$  representing the volume of traffic for class $\TrafficClass$
observed in epoch $\Epoch$.

\myparatight{Processing requirements} Each  class $\TrafficClass$ is associated with a
{\em policy service chain} or a sequence of {\em network
functions} (\NF) that process traffic in $\TrafficClass$.  Let
 $\PolicyChain = \MiddleboxFunction{1} \prec
\MiddleboxFunction{2} \prec \ldots$ denote the service
chain for class $\TrafficClass$.   These \NFs could span cellular-, IP-, and
application-level processing.  Let $\FootPrint$ denote the processing cost (e.g., CPU usage) 
 per-packet in class $\TrafficClass$ for running a  \NF 
 $\MiddleboxFunction{m}$ on a specific {\em type}  of \NF platform $\Type$. For example, 
  the per-packet CPU usage may differ 
 across virtualized vs.\ non-virtualized deployments.

\myparatight{Performance constraints} Performance constraints specify that traffic in class
$\TrafficClass$  should  have some pre-specified performance
$\ThresholdPerChain$~\cite{lte}.  As a simple starting point, we consider
the end-to-end latency for each class.

\myparatight{Cost factors} The provisioning costs associated with rolling out
the \NF platforms may depend on the platform $\Type$ (i.e., \Flexible or
\Cloud) and the location $\Location$ (e.g., power, cooling costs). We
capture these costs as follows. First, we assume that there is a {\em fixed}
cost of deploying an instance of type $\Type$ at location $\Location,
\FixedMBCost$; e.g., this captures administrative and labor costs in rolling
out new deployments and the operational costs. For deployments like \Cloud this may be zero as they may use a
pay-as-you-go model.  Second,  there is  a {\em hardware} cost,
\VariableMBCost, depending on the amount of resource provisioned for this
instance; e.g., based on the number  of CPU cores or memory  on the hardware.
Third, we have a {\em elastic} factor  depending on the actual resources used,
\ElasticMBCost; e.g., this can be linear in the amount of resources used for
\Cloud and zero for the others.

\section{Provisioning Model}

\label{sec:model}
 In this section, we describe a formal optimization framework that captures
the cost of provisioning the network to meet the time-varying processing and
performance requirements, given the policy constraints, platform costs, traffic
demands, and network specifications.   Network operators can use this framework
to (a) systematically quantify the potential benefits of NFV in their
deployments and (b) estimate the relative benefits offered by different points
in the NFV design space.

\subsection{Control Variables} 

 As a starting point, we present a model where the operator is considering  a
set of possible platform instances $\Platform$ to deploy.  As discussed in the
previous section, each $\Platform$ can be instantiated in many ways (e.g.,
\Dedicated vs.\ \Cloud) with varying cost-performance characteristics.  We use
\Type(\Platform) and \Location(\Platform) to denote the type (e.g., \Cloud or
\Dedicated) and location of a specific NF platform instance \Platform.  Now,
there are two main types of control variables that we need  to capture:

\begin{packeditemize}

\item {\em Provisioning:} The first decision is a binary decision if  we want
to deploy an instance  at a specific location. This is captured by a \{0,1\}
variable  $\ActivePlatform$.   If we choose to deploy, then we also need to
decide how much hardware resource to provision, captured by the variable
$\ResourcePerPlatform$.  For some  platform types, we also have an elastic
 option (e.g., \Flexible or \Cloud), in this case we also use 
 dynamic provisioning decision variables, $\ResourcePerPlatformPerEpoch$.

\item {\em Load distribution:} Given a provisioning strategy, we need to meet
the processing requirements and distribute the load across the various
$\Platform$ instances. In general, each class may have different chains and
the required \NFs can be instantiated at {\em any} set of  instances capable of
running these \NFs.  We can also flexibly route traffic to balance the network
and platform loads~\cite{simple}.  To capture these considerations, we introduce
{\em flow variables},   $\FractionOfTraffic$, that represents the fraction of
traffic in  class $\TrafficClass$ in epoch $\Epoch$, routed from a \NF instance
 $\MiddleboxInstance{m}$ to another \NF instance
$\MiddleboxInstanceB{m'}$ (similar to Figure~\ref{fig:flow_coverage}).  Note that we can flexibly capture different routing
strategies  by scoping these flow distribution variables differently. (Some of
these variables will not appear if $m$ or $n$ do not appear in $\PolicyChain$.)


\end{packeditemize}

\myparatight{Objective function}  Our objective is to minimize the total provisioning cost.
 This has three  components:
(1) fixed costs of instantiating platforms; (2)  hardware
 costs for  each ``active'' platform; and (3) the dynamically provisioned
compute resources per epoch.   \Eqref{eq:obj} shows this total cost in terms of the
 $\ActivePlatform$, $\ResourcePerPlatform$, and $\ResourcePerPlatformPerEpoch$
control variables:
\begin{align}
\sum \limits_{\Platform}   
&\FixedMBCostPlatform \times \ActivePlatform + \VariableMBCostPlatform \times \ResourcePerPlatform  + \nonumber \\
& \sum \limits_{\Platform,\Epoch} \ElasticMBCostPlatform \times \ResourcePerPlatformPerEpoch \label{eq:obj}
\end{align}

 As discussed earlier,  these factors depend on the type of platform; e.g., 
\Cloud may have $\FixedMBCostPlatform = \\\VariableMBCostPlatform = 0$, but
have $\ElasticMBCostPlatform>0$, while other models have  $\ElasticMBCostPlatform=0$.

\subsection{Formulation} 
Next, we describe how we capture the various processing
and provisioning constraints.

\myparatight{Resources provisioned}
 First, we need to capture the amount of resources provisioned. 
  We begin by capturing
  the total compute load on a NF instance \MiddleboxInstance{m}
 on the platform $\Platform$ during  epoch $\Epoch$ in \Eqref{eq:load:perplatperepoch}:
\begin{align}
&\forall{\Epoch, \Platform, \MiddleboxFunctionIndex}: \LoadPerFunction = 
 \nonumber \\ 
& \sum \limits_{\TrafficClass} \sum \limits_{\substack{\MiddleboxFunctionIndexGrave \text{ s.t } \MiddleboxFunctionIndexGrave \in \PolicyChain \\ \text { and }  \MiddleboxFunctionIndexGrave   \in \PlatformGrave}} \FootPrint \times \FractionOfTraffic 
 \times \TrafficVolume \label{eq:load:perplatperepoch}
\end{align}

Then, in \Eqref{eq:load:perplat}, 
the  total load on a platform $\Platform$ in an epoch $\Epoch$ is simply 
 the sum over all \NF functions that $\Platform$ can support:
\begin{equation}
\forall{\Epoch, \Platform}: \LoadPerPlatform=  \sum \limits_{\MiddleboxFunctionIndex \in \Platform}  \LoadPerFunction  \label{eq:load:perplat}
\end{equation}

Now, our provisioning strategy must ensure that 
 each \NF platform has sufficient resources to cover the
processing requirements per epoch and  across all epochs.
  Thus, we have  \Eqref{eq:res:perplatperepoch}
 and \Eqref{eq:res:perplat}:  
\begin{align}
\forall{\Epoch, \Platform}: \LoadPerPlatform = \ResourcePerPlatformPerEpoch \label{eq:res:perplatperepoch} \\
\forall{\Epoch, \Platform}: \LoadPerPlatform \leq \ResourcePerPlatform \label{eq:res:perplat}
\end{align}

In addition, and depending on other capacity constraints (e.g., space, power,
or available hardware configurations), we may also have upper bounds on the
total resources per-platform at each location:
\begin{align}
\forall{\Platform}: \ResourcePerPlatform \leq \CapPerPlatform  \label{eq:res:cap}
\end{align}

\myparatight{Fixed costs} Next, we need to model the fixed costs associated 
 with the above provisioning strategy. These fixed costs are incurred 
 if the platform is being used in {\em at least} one of the epochs 
 with non-zero resources. Thus, we have the following 
 relationship between the binary $\ActivePlatform$ 
  variables and the $\ResourcePerPlatform$ variables:

\begin{equation}
 \forall \Platform: \ResourcePerPlatform \leq \CapPerPlatform \times \ActivePlatform  \label{eq:cost:fixed}
\end{equation}

\myparatight{Modeling traffic distribution}
 The above equations model the provisioning aspects, but do not capture 
 how the traffic processing is {\em distributed} across the platforms. 
  In other words, we need to model how the $\FractionOfTraffic$ 
  variables are quantified. There are two key things we need to capture
 here.  First, we need to model the {\em coverage} constraint that 
 for each $\TrafficClass$, the desired service chain $\PolicyChain$  
 has been assigned to some set of platform instances. Second, 
 we also need to ensure that the service chain is correctly 
 applied in the intended {\em sequence}.
 Let $\PolicyChain[j]$ denote the $j^\mathit{th}$ NF in the 
 chain \PolicyChain. Note that each $\PolicyChain[j]$ may 
 be realized using several candidate platform instances. 

We model this using two sets of constraints.
 First, in \Eqref{eq:service:firsthop}  we ensure that the entire
 fraction of traffic is routed to the first hops.
\begin{equation}
\forall{\Class,\Epoch}: \sum \limits_{\MiddleboxInstance{m}:\MiddleboxFunctionIndex = \PolicyChain[1]} \FractionOfTrafficSingle = 1 \label{eq:service:firsthop}
\end{equation} 

\begin{figure}[t]
 \centering
\vspace{-0.5cm}
\includegraphics[width=220pt]{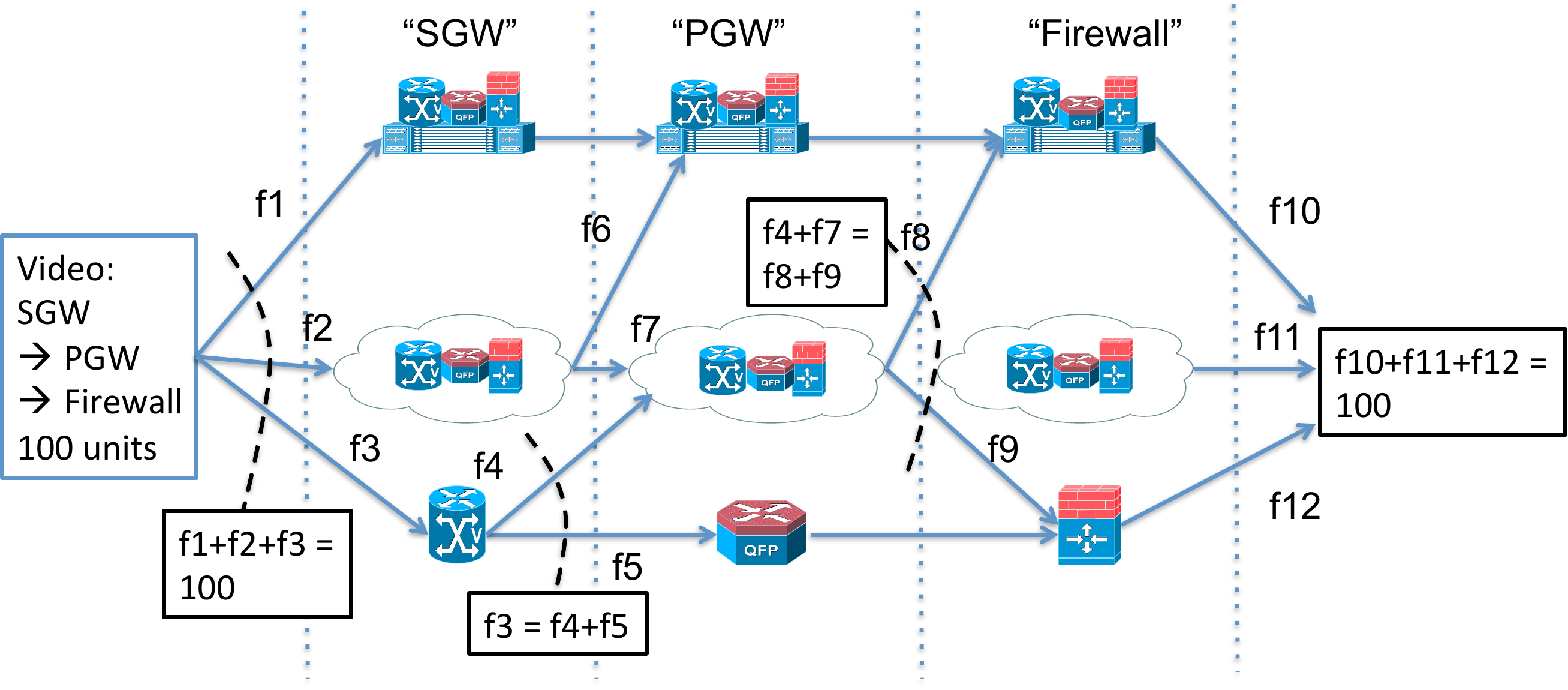}
\tightcaption{Example to illustrate how flow conservation is modeled.}
\label{fig:flow_coverage}
\end{figure}

 Second, we  model  
 {\em flow conservation} constraints that ensures that the 
 traffic incoming into one ``stage'' in the service chain 
 is routed to the next ``stage''  in the desired sequence(e.g, see Figure~\ref{fig:flow_coverage}).\footnote{
  In this paper, we are not 
 we are not mandating a specific data plane implementation. We could 
 use wildcard rules~\cite{difane} or tunnels~\cite{simple}.} 
    Then, we have: 
\begin{align}
&\forall{\MiddleboxInstance{m},\Class,\Epoch} \text{ s.t } \MiddleboxFunctionIndex =  \PolicyChain[j] \text{ \& } j>1:\nonumber \\ 
&\sum \limits_{p^{\prime}_{m'}:m' =  \PolicyChain[j-1]} \Fraction_{\Class,\Epoch,\MiddleboxInstance{\MiddleboxFunctionIndex},\MiddleboxInstanceB{\MiddleboxFunctionIndexGrave}} =  \sum \limits_{\MiddleboxInstanceB{m'} =  \PolicyChain[j+1]} \FractionOfTraffic  \label{eq:service:flowcons}
\end{align}

\myparatight{Performance bounds}
 Now, given the flow distribution variables $\FractionOfTraffic$, 
 we can also model the network-level performance that traffic 
for each class perceives. As a simple starting point, 
 we model the {\em average latency}  and ensure that this 
 is less the given threshold \ThresholdPerChain.
 If $\NetworkCostInstance$ is the typical network latency on 
  the path from $\Platform_{\MiddleboxFunctionIndex}$ to $\PlatformGrave_{\MiddleboxFunctionIndexGrave}$,
 then we can capture the performance bound as shown below:
\begin{equation}
 \forall{\Class, \Epoch}: \sum_{j=1}^{\PolicyChainSize-1}  \sum_{\substack{ \MiddleboxInstance{\MiddleboxFunctionIndex},\MiddleboxInstanceB{\MiddleboxFunctionIndexGrave} \text{s.t}  \\ \MiddleboxFunctionIndex = \PolicyChain[j] \\ \MiddleboxFunctionIndexGrave = \PolicyChain[j+1]}}
  \FractionOfTraffic  \times \NetworkCostInstance  \leq \ThresholdPerChain \label{eq:perf}
\end{equation}

\section{Example Use Cases}
\label{sec:usecase}

 Next, we highlight some illustrative use cases to validate how operators can
use our framework to evaluate  NFV design  tradeoffs.

\mypara{Setup} Due to lack of publicly available information on cellular
network topologies, we use the PoP-level  Internet2/Abilene
topology.\footnote{We have also experimented with other ISP topologies from
RocketFuel.}  We currently use four traffic classes, each with
three different \NFs. We assume there are a total of 8 different
\NFs~\cite{lte}.  We use a gravity model based on city populations as a baseline
traffic demand and simple randomized variability models. We model the different
provisioning problems as an integer linear program (ILP) and use CPLEX to solve the
problem.



\begin{figure}[h]
 \centering
\vspace{-0.5cm}
\includegraphics[width=180pt]{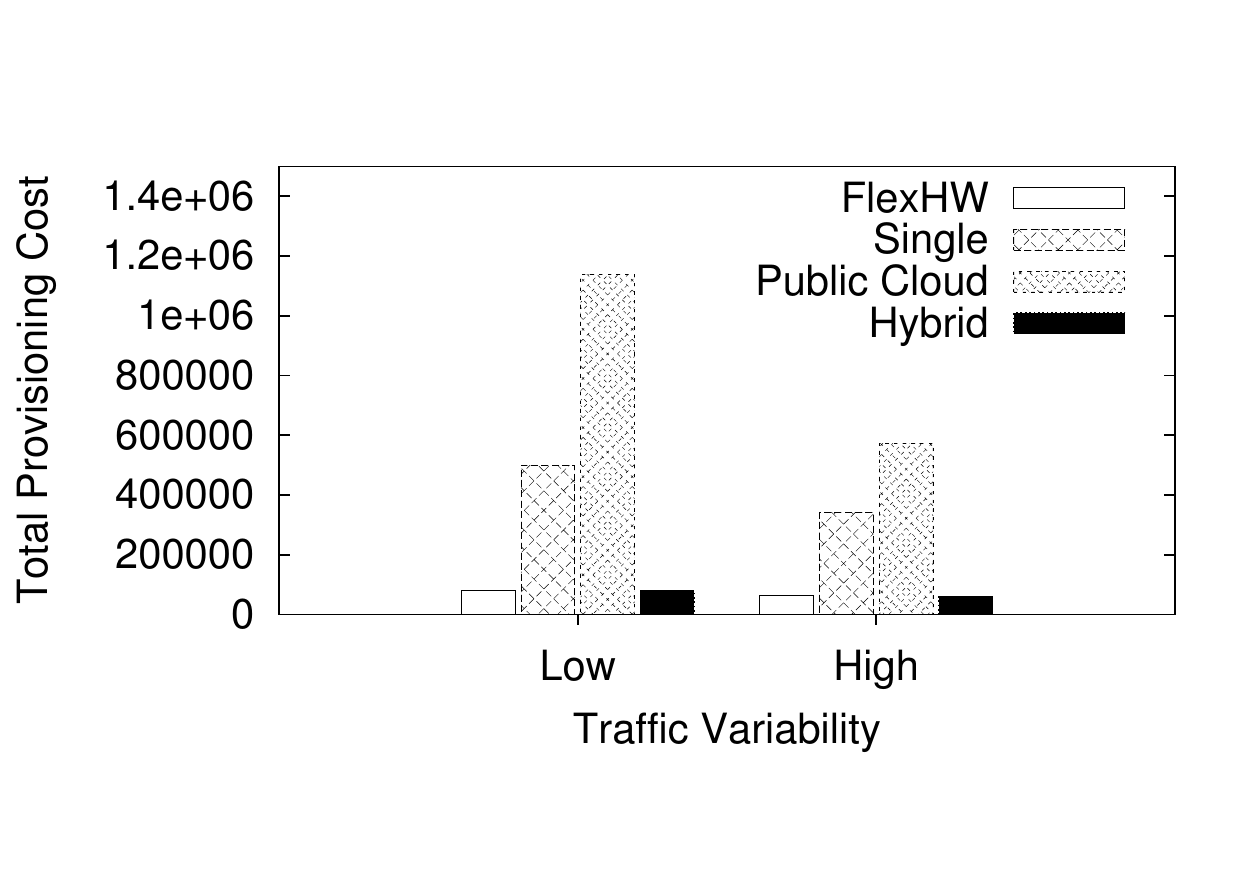}
\vspace{-0.5cm}
\tightcaption{Total provisioning cost for different NFV models.}
\label{fig:cost_dist}
\end{figure}


We instantiate the different platform types as follows: 
\begin{packeditemize}
\item \Flexible  assumes  a commodity server is being used with standard
virtualization technologies to run different functions on a single platform;
\item  \Dedicated assumes  specialized hardware for running single functions;
\item  \textit{Public Cloud} outsource  the processing to a  public cloud provider;  and 
\item  A \textit{Hybrid} deployment model which allows full flexibility to use  any combination
of the above three deployment models. 
\end{packeditemize}

 Again, given the absence of accurate cost numbers for NFV platforms, 
 we obtain ballpark numbers for the different costs by the following
strategy.  Since the different platforms have fundamentally different
cost/service models, we  normalize the costs by computing  the dollar cost  in
provisioning/running the platform for unit traffic; e.g., dollars-per-Mbps.
First, for \textit{Public Cloud}, we consider bandwidth costs in Amazon
EC2\footnote{ The CPU, memory costs were much lower than bandwidth costs}, and
we compute the normalized cost assuming  a monthly transfer volume of 500~TB.
Second, for the \Flexible hardware  we assume a typical commodity server of
price \$2,500 as representative for \Flexible. Third, for the \Dedicated
devices, we use numbers from published work and assume a specialized device at
20~Gbps capacity costs roughly \$80,000~\cite{ananta} as representative for
\Dedicated devices.  Finally, for the  setup and operational cost, we use a
common industry rule of thumb and model it to be twice the equipment
cost~\cite{opex}. 



\begin{figure*}[t]
\vspace{-1cm}
\subfloat[Varying cloud cost]
{
\vspace{-2cm}
\includegraphics[width=150pt]{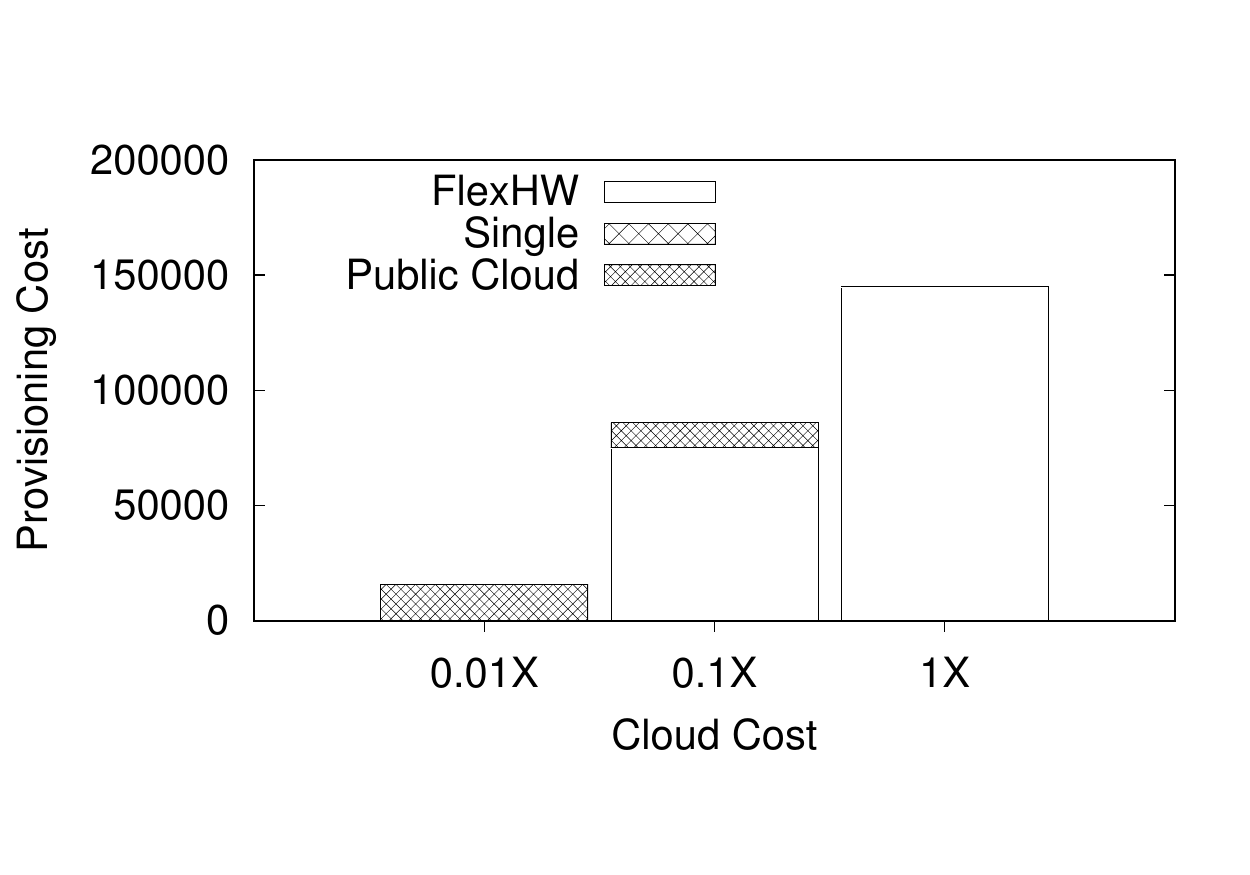}
\label{fig:cost_w_cloud}
\vspace{-2cm}
}
\subfloat[Varying setup+OPEX cost]
{
\vspace{-2cm}
\includegraphics[width=150pt]{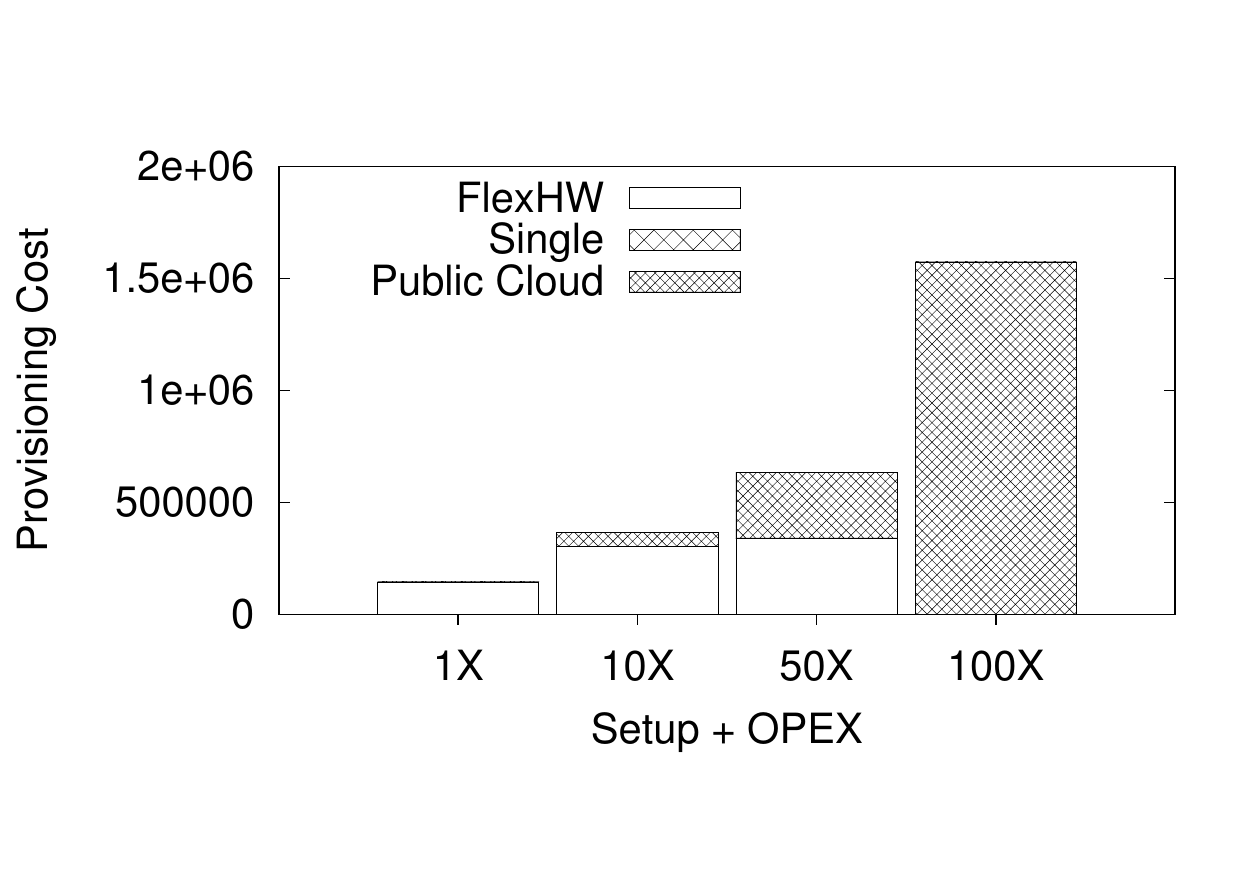}
\label{fig:cost_w_opex}
\vspace{-2cm}
}
\subfloat[Varying performance gap]
{
\vspace{-2cm}
\includegraphics[width=150pt]{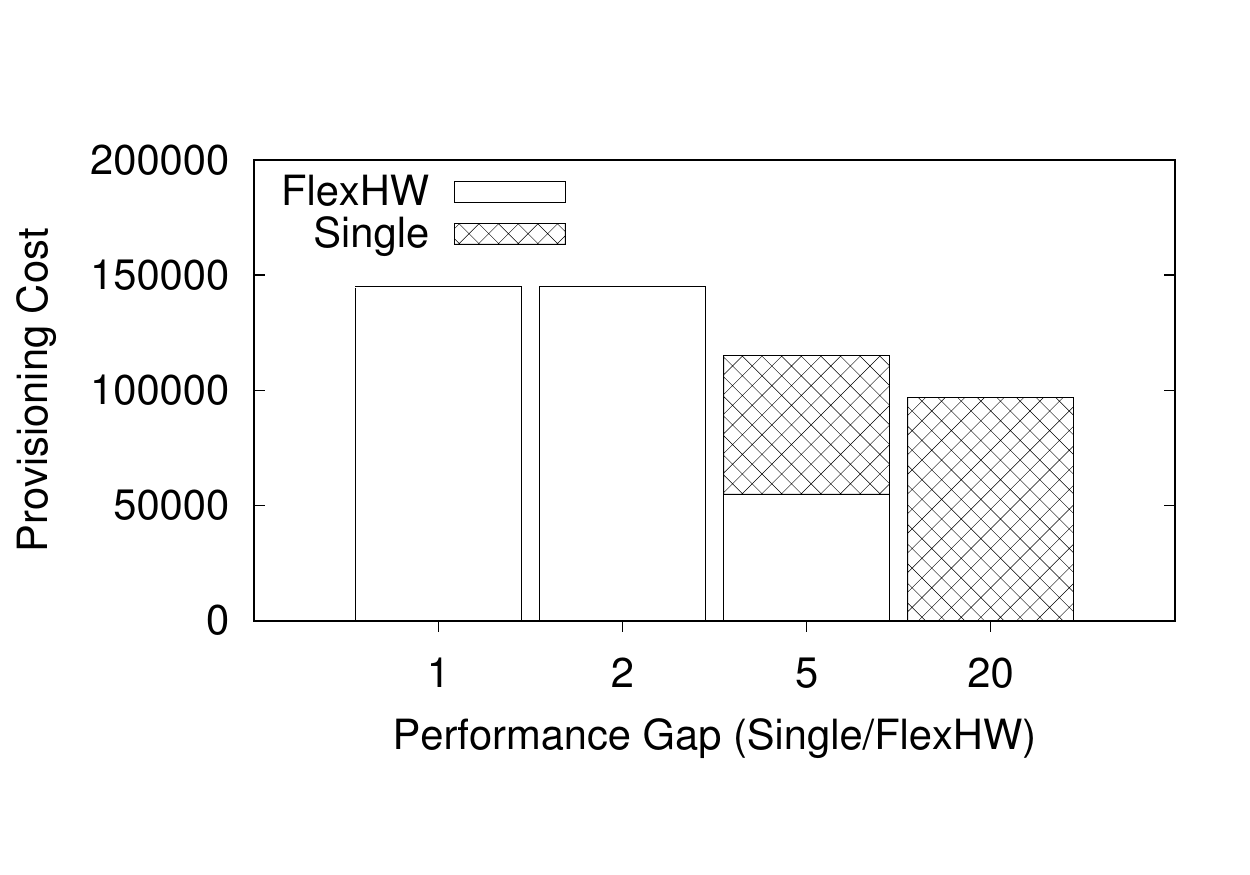}
\label{fig:cost_w_resource}
\vspace{-2cm}
}
\tightcaption{Sensitivity of the optimal strategy w.r.t. cost and performance considerations.}
\end{figure*}

\mypara{Comparing different design options}
 To illustrate this
point,   Figure~\ref{fig:cost_dist}
shows that with the given costs a deployment strategy using only \Flexible has the same provisioning cost as the \textit{Hybrid} when the traffic variability is low and very close to \textit{Hybrid} when the variability is high.




Next, we show the impact of varying different input parameters like cloud cost, setup and operational cost, and performance of each platform type. We consider a \textit{Hybrid} deployment model for these experiments with a random traffic matrix. Figure~\ref{fig:cost_w_cloud} shows that as the \Cloud cost decreases, \Cloud becomes a more viable option for processing of network functions, with a \Cloud cost of \textit{0.01X}, resulting in a \Cloud only optimal strategy. Figure~\ref{fig:cost_w_opex} shows that as the setup and operational cost increases, there is less incentive in pre-provisioning resources inside the network. Figure~\ref{fig:cost_w_resource} shows that as the performance gap between virtual appliances and specialized hardware increases, it maybe cost effective to use a combination of these two types of platforms. Note, for Figure~\ref{fig:cost_w_resource}  we only consider a \textit{Hybrid} model consisting of \Flexible and \Dedicated platforms.
\section{Related Work}
 
 To the best of our knowledge, there have been no systematic frameworks 
 to characterize the NFV design space and provide tools for
 operators to explore ``what-if'' deployment scenarios. 
 We discuss three complementary  threads of related work below. 

\myparatight{Cellular SDN and NFV} Industry reports and actual
measurements have confirmed that there are a large number of complex function
required in the cellular core~\cite{lte_whitepaper,morleysigcomm}. Today, these 
 networks are inflexible and expensive to provision/maintain. 
 This has motivated the case for network functions virtualization~\cite{nfv_whitepaper}.
 SoftCell~\cite{softcell} proposes an SDN based
 cellular core that can  support very fined-grained policies. Our
work is complementary as we focus on provisioning and placement decisions. 


\myparatight{Middleboxes} Related work  has focused on ``middlebox'' service
chaining and load balancing~\cite{simple,comb,steering}. Other work has also
suggested NFV-like ideas for traditional middleboxes~\cite{aplomb,stratos}. In
our work, we consider the provisioning and placement in addition to these
requirements.  Moreover, we present a general optimization framework that
captures all three key aspects of the NFV design space: type of platforms,
provisioning, and demand distribution. 

\myparatight{Cloud provisioning} Prior work has tried to formulate the
potential cost savings  in cloud computing~\cite{tocloud-socc11,berkeleycloud}.
As the NFV vision begins to gain momentum, we believe that there is a critical
need for frameworks like ours to shed light on the quantitative benefits that
different NFV strategies may offer.








\section{Conclusions and Future Work}
Our framework is only a first step and  we identify  several natural directions
for future work and extensions. First, our framework focuses largely on how NFV
can simplify existing deployments. We  are actively working with operators to
understand how the dynamicity and flexibility offered by NFV can enable new
deployment/service opportunities.  Second, we are also working with operators
to obtain more realistic datasets and policy requirements to quantify the
benefits of NFV in practice.   Finally, we can also extend our model to
 incorporate other kinds of policy considerations; e.g., to make sure that 
 roaming users are not co-located on the same physical hardware 
 with other users.

{
\scriptsize
\bibliographystyle{abbrv}
\bibliography{main}

\begin{thebibliography}{10}

\bibitem{opex}
{A Simple Model for Determining True Total Cost of Ownership for Data Centers}.
\newblock \url{http://tinyurl.com/kznlhn2}.

\bibitem{attdomain2.0}
{AT\&T Domain 2.0 Vision White Paper}.
\newblock \url{http://tinyurl.com/p4uv3s3}.

\bibitem{lte}
{LTE Design and Deployment Strategies}.
\newblock \url{http://tinyurl.com/lj2erpg}.

\bibitem{nfv_whitepaper}
{Network Functions Virtualisation}.
\newblock \url{http://portal.etsi.org/nfv/nfv_white_paper.pdf}.

\bibitem{lte_whitepaper}
{The LTE Network Architecture}.
\newblock \url{http://tinyurl.com/negszts}.

\bibitem{xomb}
J.~Anderson et~al.
\newblock {xOMB: Extensible Open Middleboxes with Commodity Servers.}
\newblock In {\em Proc.\ ANCS}, 2012.

\bibitem{berkeleycloud}
M.~Armbrust et~al.
\newblock Above the clouds: A berkeley view of cloud computing.
\newblock Tech Report No. UCB/EECS-2009-28.

\bibitem{tocloud-socc11}
Y.~Chen and R.~Sion.
\newblock To cloud or not to cloud? musings on costs and viability.
\newblock In {\em Proc.\ SOCC}, 2011.

\bibitem{stratos}
A.~Gember et~al.
\newblock {Stratos: A Network-Aware Orchestration Layer for Middleboxes in the
  Cloud}.
\newblock In {\em Technical Report arXiv:1305.0209}, 2013.

\bibitem{softcell}
X.~Jin et~al.
\newblock {SoftCell: Scalable and Flexible Core Network Architecture}.
\newblock In {\em CoNEXT}, 2013.

\bibitem{clickos}
J.~Martins et~al.
\newblock {Enabling fast, dynamic network processing with clickOS}.
\newblock In {\em Proc.\ HotSDN}, 2013.

\bibitem{ananta}
P.~Patel et~al.
\newblock {Ananta: Cloud Scale Load Balancing}.
\newblock In {\em Proc.\ SIGCOMM}, 2013.

\bibitem{simple}
Z.~Qazi et~al.
\newblock {SIMPLE-fying Middlebox Policy Enforcement Using SDN}.
\newblock In {\em Proc.\ SIGCOMM}, 2013.

\bibitem{comb}
V.~Sekar et~al.
\newblock {Design and Implementation of a Consolidated Middlebox Architecture}.
\newblock In {\em Proc.\ NSDI}, 2012.

\bibitem{verizononstalk}
S.Elby.
\newblock {Carrier Vision of SDN and future applications to achieve a more
  agile mobile businesss. Keynote at the OpenFlow World Congress, 2012}.

\bibitem{aplomb}
J.~Sherry et~al.
\newblock {Making Middleboxes Someone Else's Problem: Network Processing as a
  Cloud Service}.
\newblock In {\em Proc.\ SIGCOMM}, 2012.

\bibitem{nanodc}
V.~Valancius et~al.
\newblock {Greening the Internet with Nano Data Centers}.
\newblock In {\em Proc.\ CoNext}, 2009.

\bibitem{morleysigcomm}
Z.~Wang et~al.
\newblock {An Untold Story of Middleboxes in Cellular Networks}.
\newblock In {\em Proc.\ SIGCOMM}, 2011.

\bibitem{difane}
M.~Yu, J.~Rexford, M.~J. Freedman, and J.~Wang.
\newblock {Scalable Flow-Based Networking with DIFANE}.
\newblock In {\em Proc.\ SIGCOMM}, 2010.

\bibitem{steering}
Y.~Zhang et~al.
\newblock {StEERING: A Software-Defined Network for Inline Service Chaining}.
\newblock In {\em Proc.\ ICNP}, 2013.

\end{thebibliography}
}

\end{document}